\documentclass[notitlepage,superscriptaddress,showpacs,nobalancelastpage,aps,prb,twocolumn]{revtex4-1}

\RequirePackage[T1]{fontenc}
\RequirePackage{times} 

\usepackage{etoolbox}
\usepackage{amsfonts}
\usepackage{subfigure}
\usepackage{amsmath}
\usepackage{amssymb}
\usepackage{graphicx,epstopdf}
\usepackage{array}
\usepackage{color}
\usepackage{my_symbols}
\usepackage{siunitx} 
\usepackage{CJK}
\usepackage{bm}
\usepackage{tikz}
\usepackage{float}

\usepackage[italicdiff]{physics}

\usepackage{lipsum}
\usepackage[normalem]{ulem}
\usepackage{enumerate}

\usepackage[colorlinks=true, linkcolor=blue, citecolor=blue, urlcolor=blue]{hyperref}

{\par}

\makeatletter
\def\@captype{figure}
\def\figurename{Fig.}
\def\fnum@figure{\textbf{\figurename~\thefigure}}
\makeatother

\newcounter{mycomment}

\newcommand\rmv{\bgroup\markoverwith {\textcolor{red}{\rule[0.5ex]{2pt}{0.4pt}}}\ULon}

\newcommand{\Fig}[1]{{Fig. \ref{#1}}}

\begin{document}


\begin{CJK*}{UTF8}{gbsn} 

\title{Narrow waveguide based on ferroelectric domain wall}

\author{Gongzheng Chen (陈恭正)}
\affiliation{Department of Physics and State Key Laboratory of Surface Physics, Fudan University, Shanghai 200433, China}
\affiliation{Institute for Nanoelectronics Devices and Quantum Computing, Fudan University, Shanghai 200433, China}

\author{Jin Lan (兰金)}
\email[Corresponding author:~]{lanjin@tju.edu.cn}
\affiliation{Center for Joint Quantum Studies and Department of Physics, School of Science, Tianjin University, 92 Weijin Road, Tianjin 300072, China}
\affiliation{Department of Physics and State Key Laboratory of Surface Physics, Fudan University, Shanghai 200433, China}

\author{Tai Min (闵泰)}
\affiliation{Center for Spintronics and Quantum System, State Key Laboratory for Mechanical Behavior of Materials, School of Materials Science and Engineering, Xi'an Jiaotong University, Xi'an, 710049 China}

\author{Jiang Xiao (萧江)}
\email[Corresponding author:~]{xiaojiang@fudan.edu.cn}
\affiliation{Department of Physics and State Key Laboratory of Surface Physics, Fudan University, Shanghai 200433, China}
\affiliation{Institute for Nanoelectronics Devices and Quantum Computing, Fudan University, Shanghai 200433, China}
\affiliation{Shanghai Research Center for Quantum Sciences, Shanghai 201315, China}

\begin{abstract}
Ferroelectric materials are spontaneous symmetry breaking systems characterized by ordered electric polarizations.
Similar to its ferromagnetic counterpart, a ferroelectric domain wall can be regarded as a soft interface separating two different ferroelectric domains.
Here we show that two bound state excitations of electric polarization (polar wave), or the vibration and breathing modes, can be hosted and propagate within the ferroelectric domain wall.
Specially, the vibration polar wave has zero frequency gap, thus is constricted deeply inside ferroelectric domain wall, and can propagate even in the presence of local pinnings.
The ferroelectric domain wall waveguide as demonstrated here, offers new paradigm in developing ferroelectric information processing units.
\end{abstract}
 \maketitle
\end{CJK*}

A waveguide, a structure to guide wave along designated direction without loss, is one of the basic devices for wave manipulation.
Construction of waveguide lies in the heart of wave-based technologies, including optics, acoustics, magnonics, etc \cite{snyder2012optical,refi1999optical,voiculescu2012acoustic,Khitun_2010,wagner_magnetic_2016}.
Conventionally, waveguides are built upon hetero-structure consisting of different materials, which possess distinct dispersions.
As the most prominent applications, the optical fiber based on glasses of different refraction indices, has now become the standard infrastructure of the modern information society \cite{refi1999optical,kawano2004introduction}.

The downscaling of waveguide is crucial for miniaturization of the information processing devices, but is much impeded by the fabrication limit of the sharp interfaces.
An alternative approach is to make use of the existing interfaces that widely exists in ordered systems, including ferroelectric, ferromagnetic and multiferroic materials \cite{catalan2012domain}.
In ferromagnets, the magnetic domain wall, an interface between different magnetized domains, can be as narrow as tens of nanometers, due to the strong Heisenberg exchange coupling \cite{krawczyk2006magnonic,marrows2005spin}.
And recently, the magnetic domain wall is shown to act naturally as a waveguide for spin wave, in both theoretical proposals and experimental observations \cite{garcia2015narrow,lan_spin-wave_2015,wagner_magnetic_2016}.
In addition, since the magnetic domain wall is easy to create, move or eliminate, it endows the waveguide with additional flexibility \cite{beach2005dynamics,boulle2011current,klaui2003domain,parkin2008magnetic}.

In the meantime, the ferroelectric domain wall is an interface between regions of different electric polarizations, and its characteristic size is typically smaller than its magnetic counterpart \cite{meyer2002ab,seidel2012domain,catalan2012domain}.
Moreover, the collective wave-like excitation of the electric polarization, similar to spin wave, has recently attracted remarkable interests in both theoretical and experimental sides \cite{chotorlishvili_dynamics_2013,wu_low-energy_2017,yang_Domain_2020,li_Subterahertz_2021}.
Despite of close similarities with magnetic systems,  systematic investigations of the wave propagation along the ferroelectric domain wall, are only performed in limited cases \cite{wu_low-energy_2017}.






In this work, we investigate the dynamics of electric polarizations upon the background of a ferroelectric domain wall.
We show that two bound state modes, corresponding to the sliding and breathing of domain wall respectively, develop inside the domain wall.
Moreover, the propagation of these polar waves along domain wall is robust against local pinnings, as verified by numerical simulations.
The functionality of guiding polar wave, enlisting the ferroelectric domain wall a new member of domain wall waveguide family.


\emph{Model.}
Consider an one-dimensional ferroelectric wire along the $x$ direction, with its ferroelectric order pointing along $z$ direction and its strength denoted by the electric polarization $p$.
The dynamics of the electric polarization $p(x,t)$ is governed by the Landau-Khalatinikov-Tani (LKT) equation \cite{tani_dynamics_1969, ishibashi_phenomenological_1989, ishibashi_structure_1990,sivasubramanian2004physical, widom_resonance_2010, giri2011klein, chotorlishvili_dynamics_2013, khomeriki2015creation}
\begin{equation}
m \frac{{\rm d}^2p}{{\rm d} t^2} = - \gamma \frac{{\rm d} p}{{\rm d} t}+ \kappa \nabla^2 p-\frac{\partial V}{\partial p} +E,
\label{eqn:LKT}
\end{equation}
where $m$ is the effective mass, $\kappa$ is Ginzburg-type coupling constant between neighboring polarization, $V(p)$ is the Landau phenomenological potential as function of polarization $p$, $E$ applied along $z$ is the external electrical field, and $\gamma$ is the damping constant.
Rather the relaxational kinetics formulated in Ginzburg-Landau equation, the inertial dynamics in \Eq{eqn:LKT} is employed here to describe the GHz-THz dynamics of the electric polarization.

For a uniaxial ferroelectric material as considered in this work, the Landau phenomenological potential takes the minimal $p^4$ form, \ie $V(p)= -\alpha p^2/2+\beta p^4/4$. When the two phenomenological parameters $\alpha, \beta>0$ are positive, this potential leads to spontaneous ferroelectricity with two possible saturation electric polarizations $p = \pm P = \pm \sqrt{\alpha/\beta}$ in the absence of external electrical field ($E=0$).

\begin{figure}[tp]
\centering
\includegraphics[width=\columnwidth]{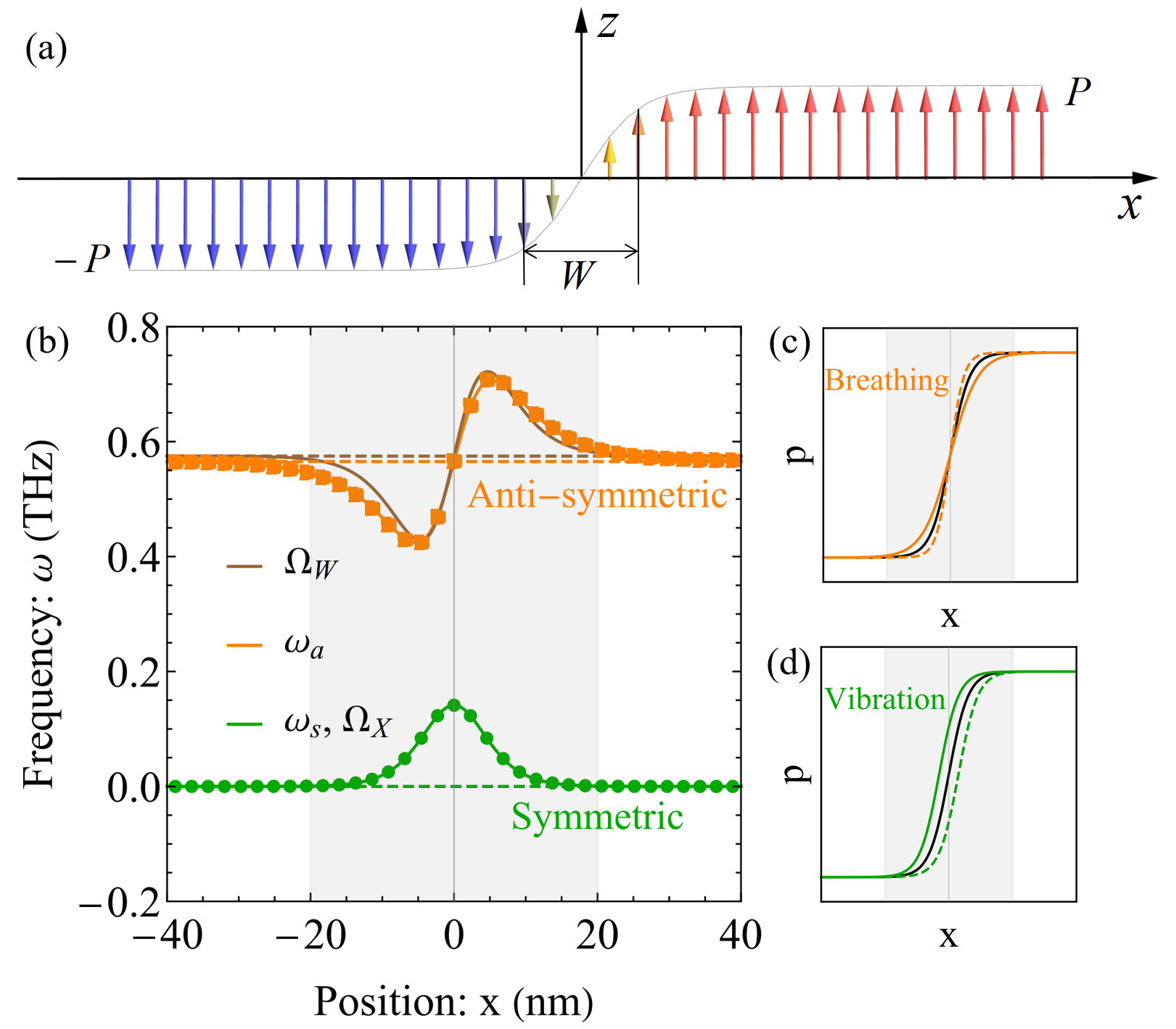}
\caption{   Ferroelectric domain wall and its bound state modes.
(a) Typical profile of the ferroelectric domain wall. The direction and length of the arrows denote the local electric polarization direction and strength, and the red/blue colors represent the up/down domains.
(b) The bound state polar waves and soft modes in the ferroelectric domain wall.
The green/orange lines plot the symmetric and anti-symmetric bound state polar waves, the corresponding frequency level is indicated by dashed lines, while results of numerical simulations are represented by dots.
(c)(d) show the total domain wall profile for the symmetric vibration mode and the anti-symmetric breathing mode.
}
\label{fig:schem_domain wall}
\end{figure}

A ferroelectric domain wall forms when two domains with different saturation polarizations meet \cite{merz1954domain}. In our case, the ferroelectric domain wall formed with two domains with $p = \pm P$ typically has a Ising-type profile \cite{ishibashi_phenomenological_1989, ishibashi_structure_1990,lee2009mixed} as depicted in \Figure{fig:schem_domain wall}(a) with its polarization strength varying as
\begin{equation}
  \label{eqn:dmw_prof}
  p_0(x) = P \tanh\frac{x-X}{W},
\end{equation}
where $X$ denotes the central position of domain wall, and the $W$ is the characteristic width. In the absence of the external electric field $E$, the characteristic width is $W_0=\sqrt{2\kappa/\alpha}$ \cite{ishibashi_phenomenological_1989, ishibashi_structure_1990}.

\emph{Polar waves within the ferroelectric domain wall.}
The magnitude of the electric polarization may also fluctuate about its equilibrium point, and leads to wave-like excitations \cite{chotorlishvili_dynamics_2013,yang_Domain_2020,li_Subterahertz_2021}, which we call {\it polar wave} here.
Separating the static and dynamical component of the polarization order parameter by $p(x,t) = p_0(x) + p'(x, t)$, where $p_0(x)$ is the static ferroelectric domain wall profile, and $p'(x,t)$ is the superimposed polar wave.
In the linear regime $\abs{p'} \ll P$, \Eq{eqn:LKT} reduces to a Klein-Gordon-like equation for $p'(x,t)$,
\begin{equation}
-m \ptt p'= \midb{- \kappa \pxx + 2\alpha + U(x) } p',
\label{eqn:LKT_dmw}
\end{equation}
where $U(x)=-3\alpha\sech^2(x/W)$ is the effective potential accounting for the inhomogeneous domain wall profile $p_0(x)$.

When the ferroelectric wire has only one homogeneous domain (no domain wall): $U = 0$, \Eq{eqn:LKT_dmw} gives the {bulk} dispersion for polar wave: $\omega_k^2 = \omega_0^2 + (\kappa/m) k^2$, with $k$ the wave-vector and $\omega_0 = \sqrt{2\alpha/m}$ the polar wave gap. When there is a domain wall, the domain wall induced potential well $U(x)$ is a special index-$2$ P\"{o}schl-Teller potential well\cite{poschl_bemerkungen_nodate}, which hosts two bound-state polar wave modes of different symmetries within domain wall. The frequencies and profiles of the symmetric and anti-symmetric bound states are
\begin{subequations}
\label{eqn:bound_state}
\begin{align}
  \mbox{sym.: }\ \omega_s &= 0, \
  &p_s' &\propto  \sech^2{x\ov W_0}, \\
  \mbox{anti-sym.: }\ \omega_a &= {\sqrt{3}\ov 2}\omega_0, \
  &p_a' &\propto   \sech{x\ov W_0}\tanh{x\ov W_0}.
\end{align}
\end{subequations}
The spatial profiles of these two modes are shown as the solid curves in \Figure{fig:schem_domain wall}(b).


\emph{Soft modes of ferroelectric domain wall.}
The bound state modes given in \Eq{eqn:bound_state} can be also regarded as the soft modes of the domain wall distortion, as shown in Fig. \ref{fig:schem_domain wall}(c)(d).
The symmetric mode, as a zero-energy Goldstone mode  costing no energy, corresponds to the vibration of the domain wall center.
While the anti-symmetric mode, with finite frequency, corresponds to the breathing mode of the domain wall with oscillating width.
These two soft modes emerging from the perturbation of position $X$ and the width $W$ have their profiles described by an effective charge distribution as
\begin{subequations}
\begin{align}
  q_\ssf{X}(x) \equiv & \pdv{p_0}{X}= -\frac{P}{W_0}\sech^2\frac{x}{W_0}, \\
  q_\ssf{W}(x) \equiv & \pdv{p_0}{W}= -\frac{P}{W_0^2} x \sech^2\frac{x}{W_0}.
\end{align}
\end{subequations}
The soft mode $q_\ssf{X}$ has symmetric charge distribution and corresponds to the vibration mode in oscillation of position $X$, and the soft mode $q_\ssf{W}$ has anti-symmetric charge distribution and corresponds to the breathing mode in oscillation of width $W$.

In the basis of soft modes $q_\ssf{X}$ and $q_\ssf{W}$, the domain wall dynamics in LKT equation \eqref{eqn:LKT} can be reduced to the dynamics of central position $X(t)$ and width $W(t)$ governed by
\begin{subequations}
\label{eqn:eom_domain wall}
\begin{align}
  \label{eqn:eom_domain wall_X}
  m_\ssf{X} \pdv[2]{X}{t} +\gamma_\ssf{X} \pdv{X}{t} &= 0 ,\\
  \label{eqn:eom_domain wall_W}
  m_\ssf{W} \pdv[2]{W}{t} +\gamma_\ssf{W} \pdv{W}{t} & = {\alpha^2\ov 3\beta} \qty({W_0^2\ov W^2}-1) \equiv R(W),
\end{align}
\end{subequations}
where $m_\ssf{X}=4mP^2/3W_0$ and $\gamma_\ssf{X}=4\gamma P^2/3W_0$ are effective mass and viscosity of the vibration mode, $m_\ssf{W}=(\pi^2-6)mP^2/9W_0$ and $\gamma_\ssf{W}=(\pi^2-6)\gamma P^2/9W_0$ are the effective mass and viscosity of the breathing mode, and $R(W)$ is the restoring force on width $W$. In \Eq{eqn:eom_domain wall}, the dynamics for domain wall position $X(t)$ and width $W(t)$ are fully decoupled, indicating that they are two independent degrees of freedom of a ferroelectric domain wall. According to \Eq{eqn:eom_domain wall_X}, the position $X$ of the domain wall is arbitrary, thus the vibration mode has zero frequency $\Omega_\ssf{X}=0$. While in \Eq{eqn:eom_domain wall_W}, the width $W$ always tends to restore to its equilibrium value $W_0$, thus the breathing mode has finite frequency $\Omega_\ssf{W}= \sqrt{ 6\alpha/m(\pi^2-6)}$.

The vibration and breathing modes of the domain wall coincide with two bound state modes discussed earlier with $\omega_s = \Omega_\ssf{X}$ and $\omega_a \simeq \Omega_\ssf{W}$. And the profile of the vibration mode $q_\ssf{X}$ is the same as the symmetric polar wave $p'_s$; while the profile for the breathing mode $q_\ssf{W}$ is also approximately the same as $p'_a$ of the anti-symmetric polar wave, as shown in \Figure{fig:schem_domain wall}(b). The agreement between domain wall soft modes and the bound state polar waves are expected, since they are the same physical excitations of the ferroelectric domain wall viewed from different perspectives. The slight deviation in frequency/profile between the domain wall breathing mode and the anti-symmetric polar wave is due to the collective coordinate description of the domain wall using only two parameters $X$ and $W$,  \ie only the domain wall position and width are allowed to vary and other distortions are forbidden.

\emph{Propagation of polar wave along ferroelectric domain wall.}
When the $1$D ferroelectric domain wall is extended to $2$D along $y$ direction, the point-like domain wall object becomes a line-shaped domain wall along $y$ direction.
Accordingly, \Eq{eqn:LKT_dmw} is modified by replacement $\partial_x^2 \ra \partial_x^2 + \partial_y^2$.
With the extra $y$ dimension, the bound state polar wave is also endowed with new freedom.
In terms of the wave vector in the $y$ direction $k_y$, the symmetric and anti-symmetric bound state polar wave modes have dispersions
\begin{equation}
  \label{eqn:disp2d}
  \omega_s(k_y) = c k_y
  \qand
  \omega_a(k_y) = \sqrt{{3\omega_0^2\ov 4} +  c^2k_y^2},
\end{equation}
where $c = \sqrt{\kappa/m}$ is the "speed of light" for the polar wave. Because the frequencies of these two modes are below the bulk gap $\omega_0$, the ferroelectric domain wall is naturally a waveguide for these two propagating polar wave modes.

Furthermore, the domain wall position and width become $y$-dependent with $X(y,t)$ and $W(y,t)$, and their dynamics are governed by
\begin{subequations}
\label{eqn:eom_domain wall_rope}
\begin{align}
\label{eqn:eom_domain wall_X2d}
m_\ssf{X} \pdv[2]{X}{t} +\gamma_\ssf{X} \pdv{X}{t} -\kappa_\ssf{X} \pdv[2]{ X}{y}&=E(X) Q , \\
\label{eqn:eom_domain wall_W2d}
m_\ssf{W} \pdv[2]{W}{t} +\gamma_\ssf{W} \pdv{W}{t} -\kappa_\ssf{W} \pdv[2]{ W}{y}&=R(W)+ E'(X) D,
\end{align}
\end{subequations}
where $\kappa_\ssf{X}=4\kappa P^2/3W_0$ and $\kappa_\ssf{W}=(\pi^2-6)\kappa P^2/9W_0$ are the effective coupling constant of vibration mode and breathing mode, and $Q=\int q_\ssf{X}(x)dx = -2P$
is the effective charge for vibration mode and $E(X)$ is the external electric field at the domain wall applied along the electric polarization direction,
and $D=\int x q_\ssf{W}(x) dx = -\pi^2 W_0P/6$ is the effective dipole for breathing mode and $E'(X)$ is the external electric field gradient acting on the dipole $D$.
\Eq{eqn:eom_domain wall_rope} defines two wave equations for waves propagating along the domain wall extending $y$-direction.
This is very much like waves on a string, with the domain wall being the string. As a consequence, the domain wall extending in $y$-direction can be regarded as a waveguide with two distinct propagating polar wave modes, whose dispersions are given by \Eq{eqn:disp2d}.


\begin{figure}[tp]
\centering
\includegraphics[width=\columnwidth]{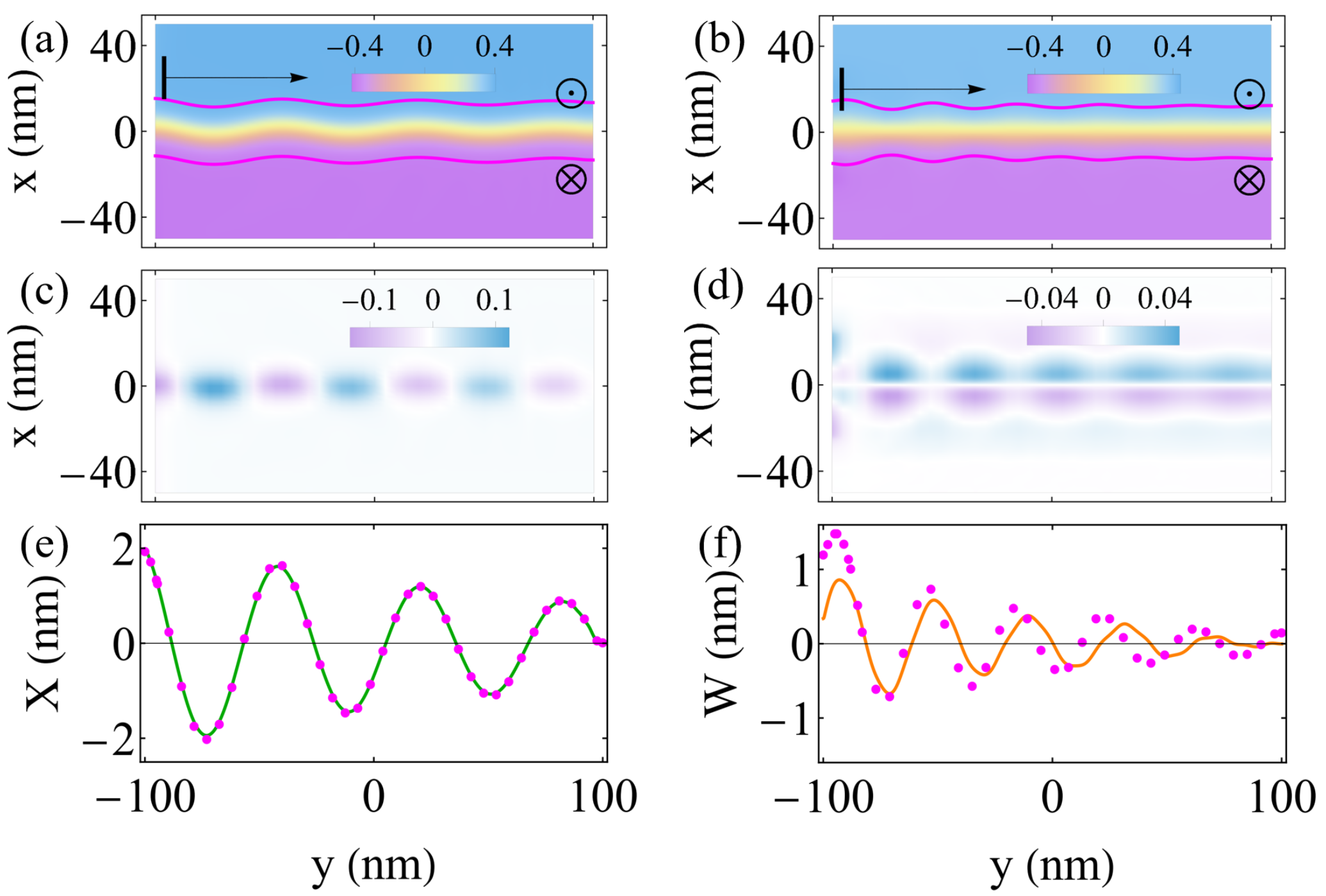}

\caption{Propagation of polar wave along the ferroelectric domain wall.
(a)(b) are the snapshots of polar wave along domain wall with the static domain wall background subtracted in a film of size $200\mathrm{nm} \times 100\mathrm{nm}$.
In (a), an oscillating electric field with frequency $\omega= \SI{0.2}{THz}$ is uniformly applied at the left side to generate symmetric polar wave; and in (b), the oscillating electric field with frequency $\omega=\SI{0.65}{THz}$ and in the form of $E(x)= K x$ is applied to generate anti-symmetric polar wave.
(c)(d) are the snapshots of the  polar wave with the electric polarization of the static domain wall subtracted corresponding to (a)(b) respectively.
(e)(f) plot the profile of domain wall position $X$ and with $W$ as functions of $y$. The solid lines are calculated from \Eq{eqn:eom_domain wall_rope}, and the dots are extracted from (a)(b).
}
\label{fig:rope}
\end{figure}

To qualitatively investigate the propagation of polar wave along ferroelectric domain wall, we perform numerical simulation based on COMSOL Multiphysics, with the time evolution model solved using the generalized alpha method. In numerical simulations, the following parameters of the ferroelectric bulk $\mathrm{BaTiO_3}$ single crystal are used \cite{chotorlishvili_dynamics_2013}:
the Landau phenomenological parameters $\alpha = \SI{2.77e7}{V\ensuremath{\cdot} m/C}$, $\beta = \SI{1.7e8}{V\ensuremath{\cdot} m^5/C^3}$, the coupling constant $\kappa = \SI{5.1e-10}{J\ensuremath{\cdot} m^3/C^2}$, the damping constant $\gamma = \SI{2.5e-5}{V\ensuremath{\cdot} m\ensuremath{\cdot} s/C}$, and the effective mass $m = \SI{1.3e-16}{V\ensuremath{\cdot} m\ensuremath{\cdot} s^2/C}$.
The bulk frequency gap for polar wave is then $\omega_0 \approx \SI{0.653}{THz}$, and the anti-symmetric mode frequency gap is $\omega_a \approx \SI{0.565}{THz}$.

In \Fig{fig:rope}(a), the symmetric polar wave with frequency $\omega = \SI{0.2}{THz}$ ($\omega<\omega_a<\omega_0$) is excited on the left edge, and the polar wave is shown to propagate freely along the domain wall.
Similarly, the anti-symmetric polar wave with frequency $\omega = \SI{0.65}{THz}$
($\omega_a<\omega<\omega_0 $) is injected in \Fig{fig:rope}(b),
and the generated polar wave is also constricted within the domain wall. The constriction of the symmetric mode is better than the anti-symmetric mode, due to its much lower frequency gap as shown in \Eq{eqn:bound_state}.

The profile of total electric polarization, including both the domain wall background and the bound state polar waves in \Fig{fig:rope}(c)(d), are demonstrated in \Fig{fig:rope}(a)(b).
Apparently, the symmetric/anti-symmetric polar wave leads to a modulation of position/width of the domain wall, as expected from the correspondence between bound state and domain wall soft modes as discussed in Sec. II. The domain wall position $X(y)$ and width $W(y)$ extracted from \Fig{fig:rope}(c)(d) are further plotted in \Fig{fig:rope}(e)(f), which agrees well with the numerical calculations based on \Eq{eqn:eom_domain wall_rope}.

\emph{Influence of impurities.}
Impurities are unavoidable in realistic materials, and are expected to affect the behavior of both the polar wave and ferroelectric domain wall  \cite{rodriguez2008ferroelectric, yang1999direct, rojac2010strong}. We model the impurity as an additional pinning potential with $V_\ssf{I} = \sum_i g p^2\delta (\br-\br_i)/2$, where $\br_i$ are the position of the impurities, and $g$ is the pinning strength. Including the impurity effect, the LKT equation is then modified to
\begin{equation}
m \frac{{\rm d}^2p}{{\rm d} t^2} = - \gamma \frac{{\rm d} p}{{\rm d} t}+ \kappa \nabla^2 p- \alpha p + \beta p^2 + E + g p\delta (\br-\br_i).
\end{equation}
Due to the locality of the impurity effect, the domain wall and polar wave basic maintain their behaviors except at these pinning sites.

%
%


Consider a ferroelectric domain wall passing through a single impurity located at $\br_0=(0,0)$. The domain wall profile is unaltered, while the polar dynamics is additionally subject to a point potential $V_\ssf{I} = gp\delta(\br)$. The scattering problem of such point-potential is complicated, therefore we turn to the domain wall distortion model in \Eq{eqn:eom_domain wall_rope}, which is modified to
\begin{equation}
  \label{eqn:X2}
m_\ssf{X} \pdv[2]{X}{t} +\gamma_\ssf{X} \pdv{X}{t} -\kappa_\ssf{X} \pdv[2]{X}{y} = E Q+ F_\ssf{PIN}(X)\delta(y),
\end{equation}
where $F_\ssf{PIN}(X)=-(gP^2/W_0) \sech^2(X/W_0)\tanh(X/W_0)$
is the pinning force acting on domain wall. Around the equilibrium position $X=0$, the pinning force is a linear restoring force $F_\ssf{PIN}(X) \approx -gP^2X/W_0^2$.
The antisymmetric breathing mode is not directly affected by the local impurity, thus is neglected in the following discussions.


Based on \Eq{eqn:X2}, we  calculate the transmission probability when the vibration mode (symmetric bound state) scatters with this impurity as:
\begin{equation}
  \label{eqn:trans_vib}
  T = \frac{1}{1+(g/g_0)^2},
\end{equation}
where $g_0=8\kappa k_y W_0/3$ is the pinning strength corresponding the transmission probability of $T=0.5$. The transmission probability is controlled by the pinning strength $g$, as well as the wave vector $k_y$ of the vibration mode.

\begin{figure}[H]
\centering
\includegraphics[width=\columnwidth]{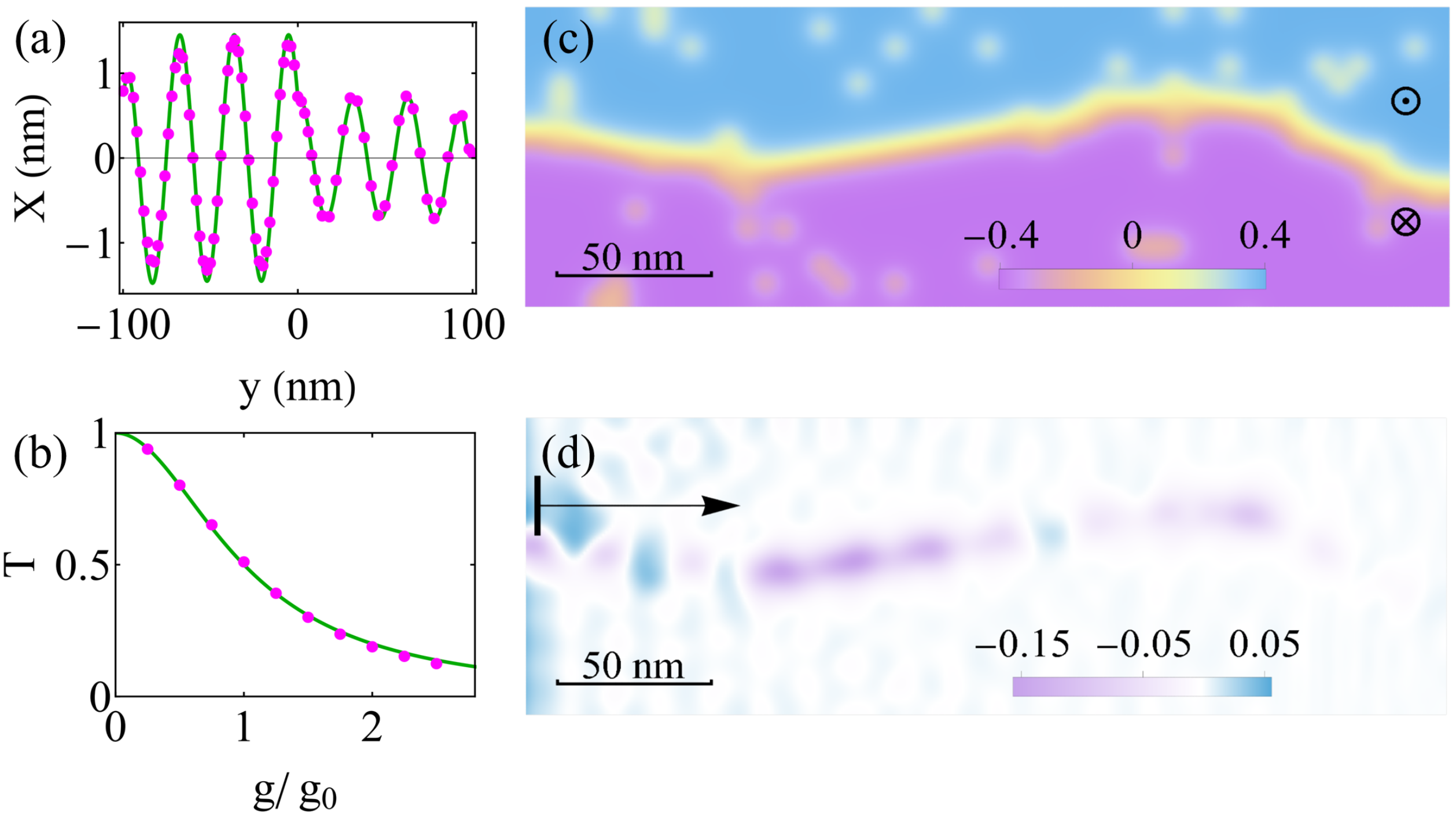}
\caption{Propagation of polar wave along the ferroelectric domain wall in the presence of impurities.
(a) Profile of the symmetric polar wave scattered by a single pinning site.
(b) The transmission probability as function of pinning strength.
The solid line is for theoretical value in \Eq{eqn:trans_vib}, and the dots are extracted from numerical simulations.
(c) The stabilized domain wall profile after relaxation. The impurities are randomly introduced inside the ferroelectric film, with averaged pinning strength of $g_0$.
(d) The snapshot of polar wave propagated along the domain wall, with the electric polarization of the static domain wall subtracted.}
\label{fig:impurity2}
\end{figure}

The transmission probability of the vibration mode along the domain wall with a single pinning site is further investigated by numerical simulations. As demonstrated in Fig.\ref{fig:impurity2}(a)(b), the transmission probability $T$ extracted from LKT equation based simulations agrees well with theoretical values in \Eq{eqn:trans_vib}. For a remarkable range of pinning strength $g$, the reflection of bound state polar wave is weak, indicating the robustness of the guiding functionality.

We proceed to investigate the influence of multiple impurities on the polar wave propagation along the domain wall. In \Fig{fig:impurity2}(c), random impurities are included in the numerical simulations, and a straight domain wall is prepared at $x=0$ for further relaxation. After relaxation, the domain wall is then captured locally by these impurities, and becomes winding as shown in \Fig{fig:impurity2}(c). An oscillating electric field with frequency $\omega = \SI{0.5}{THz} $ is then exerted at the left side of the film, thus the generated polar wave (or vibration mode) is well below the bulk frequency gap. The polar wave excited by the electric field is again highly constricted by the ferroelectric domain wall as propagation, and only experiences very little leaking and reflection. The bound state polar wave is shown in \Fig{fig:impurity2}(d), with the vibration of domain wall along the propagation clearly identified. The constricted propagation of polar wave along the winding domain wall, indicates that the self-adjusted domain wall still functions well as a waveguide.

\emph{Discussions and Conclusions.}
In this work, we focus on a $180^\circ$ up-down domain wall based on the minimal $p^4$ model of ferroelectricity, but all results naturally apply for more general types of domain wall or other ferroelectric models, such as $90^\circ$, $109^\circ$ domain wall or $p^6$ model \cite{catalan2012domain, ishibashi2002theory, rao2007domain, lubk2009first, schilling2009domains}. It is also known that the charged ferroelectric domain wall serves as channels for conduction electrons due to the modification of local chemical potential \cite{seidel2009conduction, eliseev_static_2011, farokhipoor_conduction_2011, maksymovych_dynamic_2011, schroder_conducting_2012}. In contrast, the polar wave investigate here does not involve the physical motion of electrons, thus can propagate even when the domain wall remains to be insulating.

In conclusion, we show that the ferroelectric domain wall acts as waveguide for polar wave, a collective excitation of electric polarization, similar to its magnetic counterpart. One symmetric and one antisymmetric bound state modes are identified within the ferroelectric domain wall, and they alternatively correspond to the vibration and breathing of domain wall itself. The waveguide functionality is robust again local impurities, and even survives when the shape of ferroelectric domain wall is modified substantially. The polar wave constricted within ferroelectric domain wall, offers new possibilities of transmitting electric signal in ferroelectric materials.

\emph{Acknowledgement.}
J.L. is supported by National Natural Science Foundation of China (Grant No. 11904260) and Natural Science Foundation of Tianjin (Grant No. 20JCQNJC02020). J.X. is supported by Science and Technology Commission of Shanghai Municipality (Grant No. 20JC1415900) and Shanghai Municipal Science and Technology Major Project (Grant No. 2019SHZDZX01).


%

\end{document}